\documentclass{PoS}

\newcommand{\nnu}{\nonumber\\}

\newcommand{\bef}{\begin{figure}[hbt]\centering}
\newcommand{\eef}{\end{figure}}
\newcommand{\sla}[1]{{#1}\!\!\!\slash}

\newcommand\as{\alpha_s}
\newcommand{\f}{\frac}
\newcommand{\nn}{\nonumber}
\newcommand{\GG}{{\cal G}}
\def \be  {\begin{equation}}
\def \ee  {\end{equation}}
\def \ba  {\begin{eqnarray}}
\def \ea  {\end{eqnarray}}
\def \bea  {\begin{eqnarray}}
\def \eea  {\end{eqnarray}}

\title{Semi-inclusive jet cross sections within SCET}

\ShortTitle{Semi-inclusive jet cross sections within SCET}

\author{Zhong-Bo Kang\\ 
        Theoretical Division, MS B283, Los Alamos National Laboratory, Los Alamos, NM 87545, USA\\
        E-mail: \email{zkang@lanl.gov}}

\author{\speaker{Felix Ringer}\\ 
        Theoretical Division, MS B283, Los Alamos National Laboratory, Los Alamos, NM 87545, USA\\
        E-mail: \email{f.ringer@lanl.gov}}

\author{Ivan Vitev\\ 
        Theoretical Division, MS B283, Los Alamos National Laboratory, Los Alamos, NM 87545, USA\\
        E-mail: \email{ivitev@lanl.gov}}

\abstract{We review the definition of semi-inclusive jet functions within Soft Collinear Effective Theory (SCET) and their application to inclusive jet cross sections. As an example, we consider both the inclusive production of jets and the jet fragmentation function in proton-proton collisions. The semi-inclusive jet functions satisfy renormalization group (RG) equations which take the form of standard timelike DGLAP evolution equations, analogous to collinear fragmentation functions. By solving these RG equations, the resummation of potentially large single logarithms $(\alpha_s \ln R)^n$ can be achieved. We present numerical results at NLO+NLL$_R$ accuracy and compare to existing data from the LHC.}

\FullConference{QCD Evolution 2016\\
		May 30-June 03, 2016\\
		National Institute for Subatomic Physics (Nikhef), Amsterdam}

\begin{document}

\section{Introduction}

At past and present day collider experiments, collimated jets of hadrons~\cite{Sterman:1977wj,Salam:2007xv,Cacciari:2008gp} play an important role in testing the fundamental properties of Quantum Chromodynamics (QCD), the search of physics beyond the Standard Model or for probing the quark-gluon plasma produced in heavy-ion collisions. In addition, in recent years studies of the substructure of jets have emerged as a powerful probe of QCD dynamics. Examples include the jet fragmentation function, jet broadening, jet shapes, jet mass, etc. 

Studies of jets depend on the choice of jet definition and the jet radius parameter $R$. This parameter serves as a distance measure between particles and determines when particles are clustered together in the same jet. For inclusive jet observables, the jet radius parameter shows up in the perturbative expansion as single logarithms $(\as\ln R)^n$~\cite{Jager:2004jh,Mukherjee:2012uz,deFlorian:2013qia}. For narrow jets, i.e. small values of $R$, these logarithms can be large and need to be resummed to all orders in perturbation theory~\cite{Dasgupta:2014yra,Dasgupta:2016bnd}. This can be achieved with the framework presented in this work. In~\cite{Kang:2016mcy,Kang:2016ehg} the so-called ``semi-inclusive jet functions'' were first introduced and applied to the calculation of jet cross sections. Their exact definition will be given below. In this work, we consider both the inclusive production of jets as well as the jet fragmentation function (JFF). The JFF describes the longitudinal momentum distribution of hadrons inside a fully reconstructed jet~\cite{Procura:2009vm,Jain:2011xz,Arleo:2013tya,Kaufmann:2015hma,Chien:2015ctp,Dai:2016hzf}. It is one of the phenomenologically most relevant jet substructure observables~\cite{Chatrchyan:2012gw,Aad:2013ysa,ATLAS:2015mla}. For inclusive jet production and for the JFF, we consider jets that are produced in proton-proton collisions. We rely on the framework of Soft Collinear Effective Theory (SCET)~\cite{Bauer:2000yr,Bauer:2001yt,Beneke:2002ph}. In past years, SCET has emerged as a valuable tool to describe various jet observables. A resummation of small-$R$ logarithms is highly desired since small values of $R$ are frequently used in jet observables by the experiments, especially for jet substructure. Relatively small jet radii, as small as $R=0.2$, are also commonly used in heavy ion collisions~\cite{Aad:2012vca,Abelev:2013kqa}, in order to suppress the heavy-ion background.

In this paper, within the framework of SCET, we review the definition of the semi-inclusive jet functions $J_i(z, \omega_J, \mu)$ that describe a jet with energy $\omega_J$ and radius $R$ carrying a fraction $z$ of the light cone energy of the parton $i$ that initiates the jet~\cite{Kang:2016mcy}. This new kind of jet function can be used to calculate inclusive jet cross sections. We present results for $J_i(z, \omega_J, \mu)$ up to next-to-leading order (NLO) and its renormalization group (RG) equations. The resulting RG equations take the form of standard timelike DGLAP equations, satisfied also by collinear fragmentation functions. By working out their solution, we are able to resum single logarithms in the jet radius parameter $(\alpha_s\ln R)^n$ at next-to-leading logarithmic (NLL$_R$) accuracy. We present a factorization theorem for the process $pp\to {\rm jet}X$ which may be written in terms of hard-scattering functions for inclusive hadron production~\cite{Jager:2002xm} and the evolved semi-inclusive jet function. Analogously, we introduce the semi-inclusive fragmenting jet function (FJF) $\GG_i^h(z,z_h,\omega_J,R,\mu)$~\cite{Kang:2016ehg}. The only additional variable, $z_h$, is given by the ratio of the longitudinal momentum components of the hadron and the jet. The factorized form of the cross section for $pp\to ({\rm jet}\, h)X$ is essentially the same as for $pp\to\mathrm{jet}X$ but the semi-inclusive jet function is replaced with the semi-inclusive FJF. Finally, we present numerical results for both observables for LHC kinematics at a combined accuracy of NLO+NLL$_R$.

The remainder of this paper is organized as follows. In Sec.~\ref{sec:two}, we give the definition of the semi-inclusive jet function. In addition, we present its RG equations and outline their solution. We present a factorized form of the cross section in terms of hard-scattering functions and the semi-inclusive jet functions for the process $pp\to {\rm jet}X$ and present numerical results for the LHC. In Sec.~\ref{sec:three}, we proceed by introducing the semi-inclusive FJF. We present numerical results for the process $pp\to ({\rm jet}\, h)X$ and compare to available data from the LHC. We conclude our paper in Sec.~\ref{sec:four}.

\section{Inclusive jet production\label{sec:two}}

\bef
\includegraphics[width=4.2in]{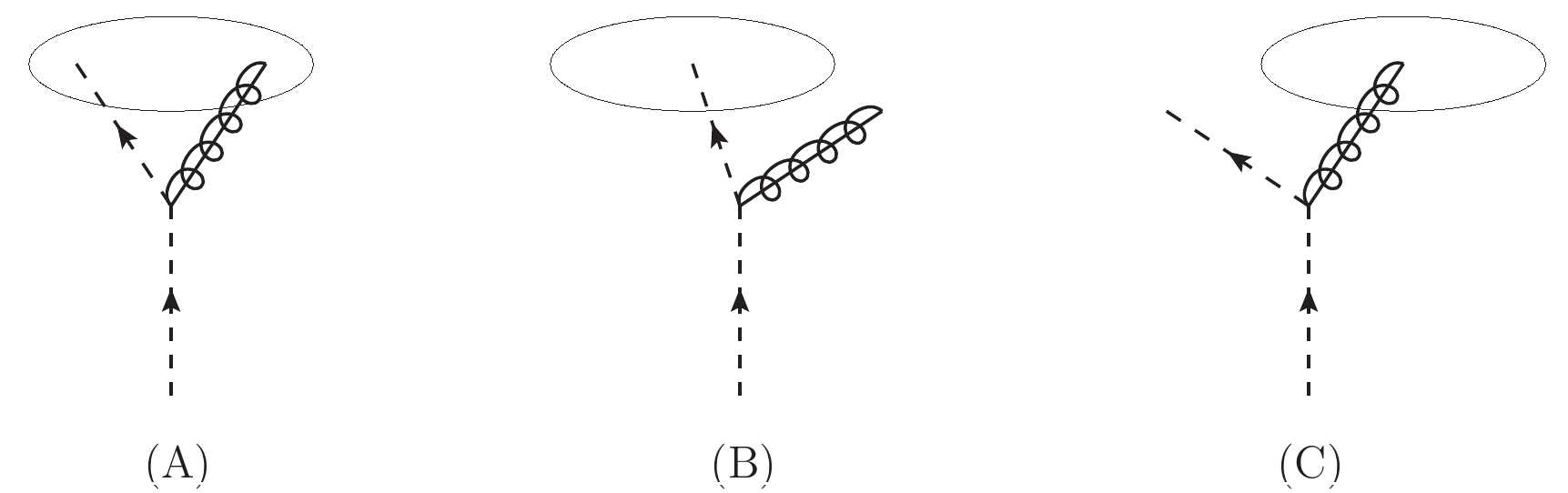}
\caption{Three situations that contribute to the semi-inclusive quark jet function: (A) both tha quark and tha gluon are inside the jet, (B) only the quark is inside the jet, (C) only the gluon is inside the jet. The dashed (curly-solid) line represents a collinear quark (gluon). \label{fig:configuration}}
\eef

We start by presenting the definition of the semi-inclusive jet functions for  quarks and gluons $J_q(z, \omega_J)$ and $J_g(z, \omega_J)$, respectively~\cite{Kang:2016mcy}
\ba
J_q(z = \omega_J / \omega, \omega_J, \mu) &= \frac{z}{2N_c}{\rm Tr} \left[\frac{\sla{\bar n}}{2}
\langle 0| \chi_n(0) \delta\left(\omega - \bar n\cdot {\mathcal P} \right) |JX\rangle \langle JX|\bar \chi_n(0) |0\rangle \right],
\\
J_g(z = \omega_J / \omega, \omega_J, \mu) &= \frac{z\,\omega}{2(N_c^2-1)}
\langle 0| {\mathcal B}_{n\perp \mu}(0) 
 \delta\left(\omega - \bar n\cdot {\mathcal P} \right) |JX\rangle \langle JX|{\mathcal B}_{n\perp}^\mu(0)  |0\rangle.
\ea
Here, $\omega$ is the energy of the initiating parton $i$, ${\cal P}$ is the label momentum operator and $\chi_n$, ${\mathcal B}_{n\perp}^\mu$ are gauge invariant building blocks for quark and gluons fields in SCET, respectively. Furthermore, $|JX\rangle$ represents the final state, where $J$ is the observed jet and $X$ the unobserved hadronic remnant. See~\cite{Kang:2016mcy} for more details. To leading-order, we simply have $J_{q,g}^{(0)}(z,\omega_J)=\delta(1-z)$. As an example, we present the NLO result for the semi-inclusive jet function for an initial quark. To NLO, we need to consider the diagrams  shown in Fig.~\ref{fig:configuration}. All three diagrams show the same $q\to qg$ branching process but we need to distinguish the situation A) where both partons remain inside the jet and B), C) where one parton exits the jet. For example, for the anti-$k_T$ algorithm at order ${\cal O}(\as)$, we find the following result after renormalization
\ba
J_q^{(1)}(z, \omega_J)  & = & \frac{\alpha_s}{2\pi} \ln\left(\f{\mu^2}{\omega_J^2\tan^2(R/2)}\right) 
\Big[P_{qq}(z) + P_{gq}(z)\Big] -\frac{\alpha_s}{2\pi} \Bigg\{ C_F\bigg[2\left(1+z^2\right)\left(\frac{\ln(1-z)}{1-z}\right)_{+} 
\nnu
&&
+ (1-z) \bigg] 
- \delta(1-z)\,C_F \left(\f{13}{2}-\f{2\pi^2}{3}\right) +P_{gq}(z) 2\ln\left(1-z\right) + C_F z \Bigg\}\,.
\label{eq:J1q}
\ea
Here, $P_{qq}(z)$ and $P_{gq}(z)$ are the standard Altarelli-Parisi splitting functions. Note that we are left with only a single logarithmic dependence on the jet parameter $R$, here $\ln\left(\mu^2/(\omega_J^2\tan^2(R/2))\right)$, which we eventually want to resum. For exclusive jet observables a double logarithmic dependence is obtained, see e.g.~\cite{Ellis:2010rwa,Kolodrubetz:2016dzb}. Due to renormalization, the semi-inclusive jet functions $J_{q/g}(z, \omega_J, \mu)$ satisfy the following RG equation
\bea
\mu \frac{d}{d\mu} J_i(z, \omega_J, \mu) = \frac{\alpha_s(\mu)}{\pi} \sum_j \int_z^1  \frac{dz'}{z'} P_{ji}\left(\frac{z}{z'}, \mu \right) J_j(z', \omega_J, \mu).
\label{eq:evo}
\eea
The anomalous dimensions that appear here on the right hand side, $P_{ji}(z)$, are again the usual Altarelli-Parisi splitting functions. Therefore, the RG equation is exactly the same as the timelike DGLAP equations satisfied by collinear fragmentation functions, $D_i^h(z, \mu)$, which describe the non-perturbative fragmentation of a parton $i$ into a hadron $h$. The fact that we obtain DGLAP-type evolution equations is a characteristic feature of inclusive jet observables. In contrast, for exclusive jet observables multiplicative RG equations are obtained, see e.g.~\cite{Ellis:2010rwa,Kolodrubetz:2016dzb}. This new type of RG equations for semi-inclusive jet functions in Eq.~(\ref{eq:evo}), necessary for inclusive jet cross sections, constitutes the main new result of our work. 
\bef
\includegraphics[width=3.5in,trim=0cm 0cm 0cm -0.5cm]{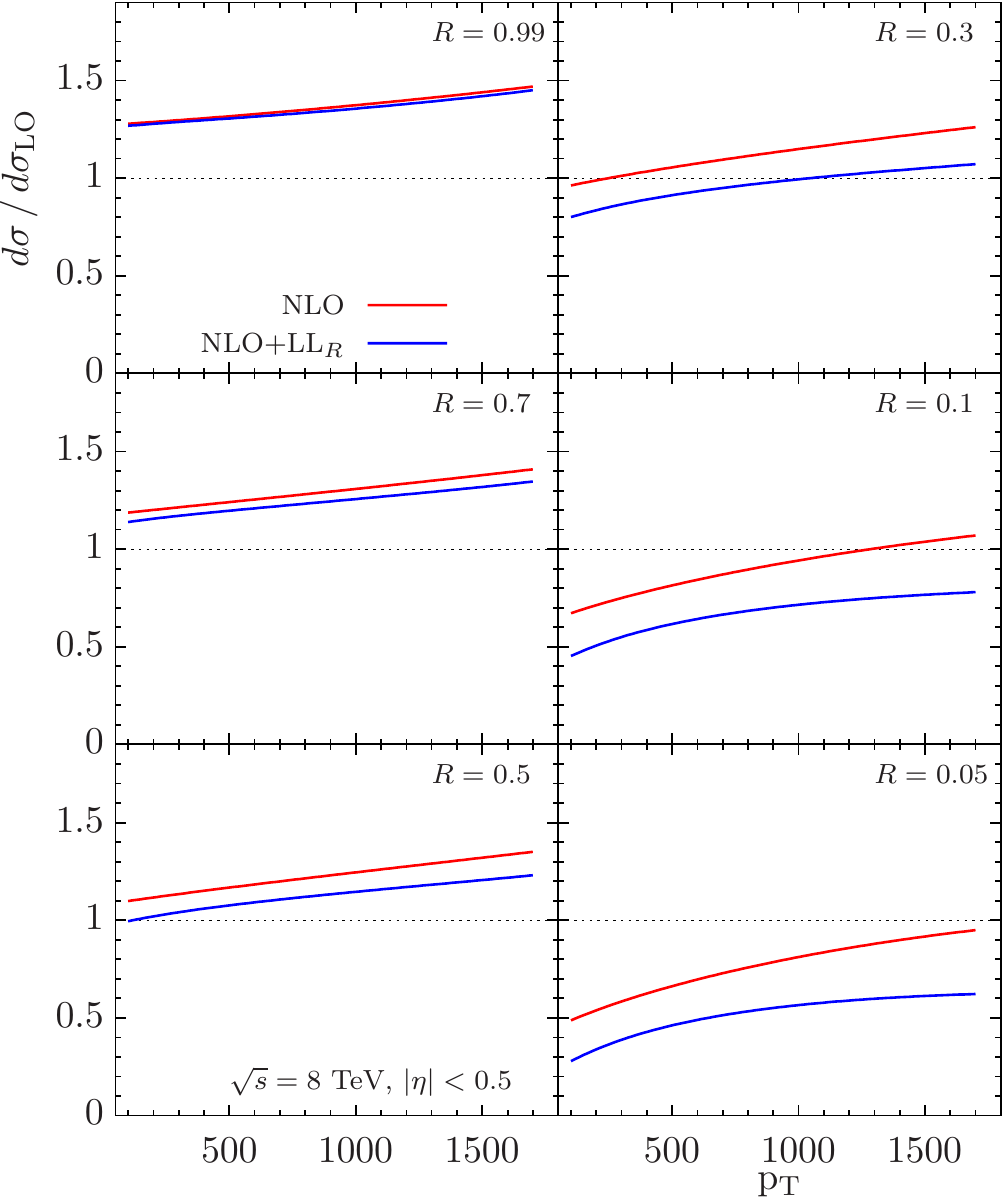}
\caption{NLO (red) and ${\rm NLO+LL}_R$ (blue) cross sections normalized to the leading-order result for different values of $R=0.99-0.05$. The small-$R$ approximation is only up to $R=0.7$. As an example, we choose $\sqrt{s}=8$~TeV and $|\eta|<0.5$. \label{fig:cross-res}}
\eef
From the NLO expressions in Eq.~(\ref{eq:J1q}), it can be seen that the natural scale for the semi-inclusive jet function is given by $\mu \sim \omega_J \tan(R/2)$ for which the large logarithmic terms are eliminated. Thus, by solving the evolution equations in Eq.~(\ref{eq:evo}) above from the scale $\mu_J \sim \omega_J \tan(R/2)$ to a higher scale $\mu\sim p_T$, we resum logarithms of the form $(\as \ln R )^n$. Following the methods developed in~\cite{Vogt:2004ns,Anderle:2015lqa,Bodwin:2015iua}, we can work out the solution of the DGLAP-type evolution equations in Mellin space at NLL$_R$. See~\cite{Kang:2016mcy} for more details.

We proceed by presenting a factorization theorem for inclusive jet production in proton-proton collisions. Following~\cite{Jager:2004jh,Kang:2016mcy,Kaufmann:2015hma,Jager:2002xm,Fickinger:2016rfd}, one finds that the cross section for $pp\to\mathrm{jet}X$ can be cast into the following form
\be
\label{eq:sigjetX}
\frac{d\sigma^{pp\to {\rm jet}X}}{dp_Td\eta}  = \frac{2 p_T}{s}\sum_{a,b,c}\int_{x_a^{\rm min}}^1\f{dx_a}{x_a}f_a(x_a,\mu)\int_{x_b^{\rm min}}^1\f{dx_b}{x_b} f_b(x_b,\mu) \int^1_{z_c^{\rm min}} \frac{dz_c}{z_c^2}\frac{d\hat\sigma^c_{ab}(\hat s,\hat p_T,\hat \eta,\mu)}{dvdz}J_c(z_c,\omega_J,\mu)\,.
\ee
Here, $J_c$ is the evolved semi-inclusive jet function, $f_{a,b}$ are the parton distribution functions (PDFs) for the initial protons and the $d\hat\sigma_{ab}^c/dvdz$ are the hard functions. As demonstrated in~\cite{Kang:2016mcy}, the hard functions here are the same as for inclusive hadron production~\cite{Jager:2002xm}. In addition, $p_T$ and $\eta$ denote the transverse momentum and the rapidity of the observed final state jet respectively. The definition of all other kinematic variables and the lower integration limits in Eq.~(\ref{eq:sigjetX}) can be found in~\cite{Kang:2016mcy}. Note that an analogous version of Eq.~(\ref{eq:sigjetX}) was already formulated in~\cite{Kaufmann:2015hma} using only the fixed order results for the semi-inclusive jet function as in Eq.~(\ref{eq:J1q}). 

Finally, in Fig.~\ref{fig:cross-res}, we present numerical results showing the phenomenological  relevance of $\ln R$ resummation for LHC kinematics. We show the NLO (red) and ${\rm NLO+LL}_R$ (blue) cross sections normalized to the leading-order result for different values of $R=0.99-0.05$. We choose $\sqrt{s}=8$~TeV and $|\eta|<0.5$ and, throughout this work, we use the PDFs of~\cite{Nadolsky:2008zw}. The difference between the NLO calculation and the $\ln R$ resummed result can be as large as $30\%$ for $R=0.1$ and $p_T=1700$ GeV.

\section{The jet fragmentation function\label{sec:three}}

\bef
\includegraphics[width=3.6in,trim=0cm 2cm 0cm 2cm]{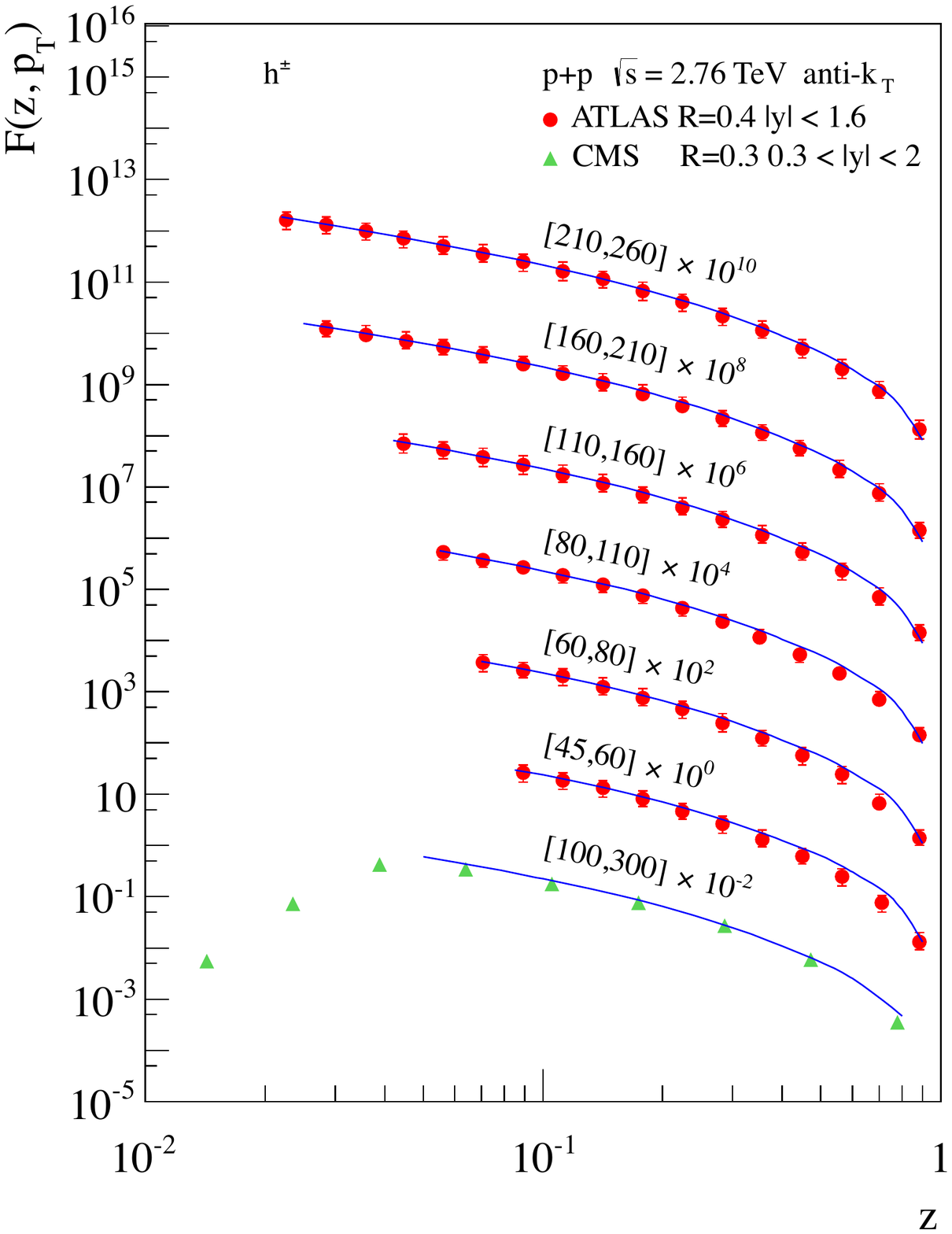}
\caption{Comparison of our numerical calculations (solid blue lines) to  LHC data in proton-proton collisions at $\sqrt{s} = 2.76$ TeV. The solid red circles correspond to the preliminary ATLAS data  form Ref.~\cite{ATLAS:2015mla} and the green triangles are the CMS data from Ref.~\cite{Chatrchyan:2012gw}. The numbers in the square brackets correspond to different jet transverse momentum bins in the range of $45-300$~GeV.}
\label{fig:ATLASCMS}
\eef

In order to describe the jet fragmentation function, where a hadron is identified inside a reconstructed jet, we need to introduce a different kind of semi-inclusive jet function. In analogy to the derivation outlined in section~\ref{sec:two}, we define the semi-inclusive fragmenting jet function for quarks and gluons as
\ba
\GG_q^h(z, z_h, \omega_J, \mu) &=& \frac{z}{2N_c} \delta\left(z_h - \frac{\omega_h}{\omega_J}\right)
{\rm Tr} \left[\frac{\sla{\bar n}}{2}
\langle 0| \delta\left(\omega - \bar n\cdot {\mathcal P} \right) \chi_n(0)  |(Jh)X\rangle \langle (Jh)X|\bar \chi_n(0) |0\rangle \right],
\nn \\Ê\\
\GG_g^h(z, z_h, \omega_J, \mu) &=& - \frac{z\,\omega}{(d-2)(N_c^2-1)} \delta\left(z_h - \frac{\omega_h}{\omega_J}\right)
\langle 0|  \delta\left(\omega - \bar n\cdot {\mathcal P} \right) {\mathcal B}_{n\perp \mu}(0) 
 |(Jh)X\rangle 
 \nnu
&&\hspace{50mm} \times \langle (Jh)X|{\mathcal B}_{n\perp}^\mu(0)  |0\rangle\, .
\ea
The additional variable $z_h$ is given by the ratio of the longitudinal momenta of the hadron and the jet, as mentioned above. When evaluating the semi-inclusive FJF at NLO, we need to consider again the diagrams shown in Fig.~\ref{fig:configuration}. To this order in perturbation theory, we find that either $z=1$ or $z_h=1$. See~\cite{Kang:2016ehg} for a detailed calculation. The renormalization of UV poles and the derivation of an analogous RG equation as in~(\ref{eq:evo}) follows the same steps as outlined in section~\ref{sec:two}. The resulting DGLAP-type RG equation can be used again to resum single logarithms in the jet parameter $R$. The remaining IR poles are matched onto collinear fragmentation functions as
\bea
\label{eq:matching}
\GG_i^h(z,z_h,\omega_J,\mu) = \sum_j \int_{z_h}^1 \frac{dz_h'}{z_h'} {\mathcal J}_{ij}\left(z,z_h',\omega_J,\mu\right) D_j^h\left(\frac{z_h}{z_h'},\mu\right) \, .
\eea
The final results for the matching coefficients ${\cal J}_{ij}$ and a factorized form of the cross section for proton-proton scattering similar to~(\ref{eq:sigjetX}) can be found in~\cite{Kang:2016ehg}. See also~\cite{Kaufmann:2015hma}. We present numerical results for the jet fragmentation function which is given by the ratio
\bea
\label{eq:JFFdef}
F(z_h,p_T)=\frac{d\sigma^{pp\to (\mathrm{jet}h)X}}{dp_Td\eta dz_h}\Big/ \frac{d\sigma^{pp\to \mathrm{jet}X}}{dp_Td\eta},
\eea
where $z_h=p_T^h/p_T$, $p_T$ and $\eta$ are integrated over certain bins. In Fig.~\ref{fig:ATLASCMS}, we show our numerical results at NLO+NLL$_R$ accuracy for $\sqrt{s}=2.76$~TeV using the fragmentation functions of~\cite{deFlorian:2007ekg} and compare to data from ATLAS~\cite{ATLAS:2015mla} and CMS~\cite{Chatrchyan:2012gw}. The numbers in the square brackets correspond to different jet transverse momentum bins in the range of $45-300$~GeV. We find very good agreement between theory and data.

\section{Conclusions\label{sec:four}}

In this work, we introduced two new kinds of jet functions within SCET, relevant for the evaluation of inclusive jet observables. First, we considered the semi-inclusive jet function, which appears in the factorized formalism for single inclusive jet production. We calculated the new jet function to NLO and derived the corresponding RG equations which take the form of timelike DGLAP evolution equations, satisfied also by collinear fragmentation functions. By solving these DGLAP-type equations, we were able to resum single logarithms in the jet radius parameter $(\as \ln R)^n$ to NLL$_R$ accuracy. We found numerically significant effects, e.g. $\ln R$ resummation can lead to a reduction of $10-20\%$ of the cross section compared with the NLO results for intermediate $R=0.3-0.5$. Secondly, we considered the semi-inclusive fragmenting jet function which is needed to describe the jet fragmentation function where a hadron is identified inside a reconstructed jet. In direct analogy to inclusive jet production, we obtained DGLAP type evolution equations which allow for the resummation of single logarithms in the jet radius parameter $R$. We compared our numerical results to existing data from the LHC and found very good agreement. Our results are also applicable to $e^+e^-$ annihilation  and $ep$ scattering, the latter being particularly important for a future Electron-Ion Collider (EIC). We plan to extend our new formalism in the future to other jet substructure observables, which can now be calculated as inclusive quantities. In addition, we expect that this framework will facilitate the combination of $\ln R$ resummation with other types of resummation, such as threshold resummation. Finally, we expect significant improvements from our new framework in the ability to describe jet substructure observables in heavy-ion collisions~\cite{Chien:2015hda,Chien:2016led} in a suitably modified effective theory~\cite{Ovanesyan:2011kn}.

\section*{Acknowledgments} 

We would like to thank Patriz Hinderer, Christopher Lee, Yan-Qing Ma, Emanuele Mereghetti, Asmita Mukherjee, Piotr Pietrulewicz, Tom Kaufmann, Ignazio Scimemi, Iain Stewart, Frank Tackmann, Werner Vogelsang, and Wouter Waalewijn for very helpful discussions and useful comments. This work is supported by the U.S. Department of Energy under Contract No.~DE-AC52-06NA25396, and in part by the LDRD program at Los Alamos National Laboratory.


\begin{thebibliography}{99}

\bibitem{Sterman:1977wj} 
  G.~F.~Sterman and S.~Weinberg,
  Phys.\ Rev.\ Lett.\  {\bf 39}, 1436 (1977),

\bibitem{Salam:2007xv} 
  G.~P.~Salam and G.~Soyez,
  JHEP {\bf 0705}, 086 (2007),
  arXiv:0704.0292 [hep-ph].

\bibitem{Cacciari:2008gp} 
  M.~Cacciari, G.~P.~Salam and G.~Soyez,
  JHEP {\bf 0804}, 063 (2008),
  arXiv:0802.1189 [hep-ph].

\bibitem{Jager:2004jh} 
  B.~Jager, M.~Stratmann and W.~Vogelsang,
  Phys.\ Rev.\ D {\bf 70}, 034010 (2004),
 [hep-ph/0404057].

\bibitem{Mukherjee:2012uz} 
  A.~Mukherjee and W.~Vogelsang,
  Phys.\ Rev.\ D {\bf 86}, 094009 (2012)
  arXiv:1209.1785 [hep-ph].

\bibitem{deFlorian:2013qia} 
  D.~de Florian, P.~Hinderer, A.~Mukherjee, F.~Ringer and W.~Vogelsang,
  Phys.\ Rev.\ Lett.\  {\bf 112}, 082001 (2014),
  arXiv:1310.7192 [hep-ph].

\bibitem{Dasgupta:2014yra} 
  M.~Dasgupta, F.~Dreyer, G.~P.~Salam and G.~Soyez,
  JHEP {\bf 1504}, 039 (2015)
  arXiv:1411.5182 [hep-ph].

\bibitem{Dasgupta:2016bnd} 
  M.~Dasgupta, F.~A.~Dreyer, G.~P.~Salam and G.~Soyez,
  JHEP {\bf 1606}, 057 (2016)
  arXiv:1602.01110 [hep-ph].

\bibitem{Kang:2016mcy} 
  Z.~B.~Kang, F.~Ringer and I.~Vitev,
  arXiv:1606.06732 [hep-ph].

\bibitem{Kang:2016ehg} 
  Z.~B.~Kang, F.~Ringer and I.~Vitev,
  arXiv:1606.07063 [hep-ph].

\bibitem{Procura:2009vm} 
  M.~Procura and I.~W.~Stewart,
  Phys.\ Rev.\ D {\bf 81}, 074009 (2010)
  Erratum: [Phys.\ Rev.\ D {\bf 83}, 039902 (2011)],
  arXiv:0911.4980 [hep-ph].

\bibitem{Jain:2011xz} 
  A.~Jain, M.~Procura and W.~J.~Waalewijn,
  JHEP {\bf 1105}, 035 (2011),
  arXiv:1101.4953 [hep-ph].

\bibitem{Arleo:2013tya} 
  F.~Arleo, M.~Fontannaz, J.~P.~Guillet and C.~L.~Nguyen,
  JHEP {\bf 1404}, 147 (2014),
  arXiv:1311.7356 [hep-ph].

\bibitem{Kaufmann:2015hma} 
  T.~Kaufmann, A.~Mukherjee and W.~Vogelsang,
  Phys.\ Rev.\ D {\bf 92}, no. 5, 054015 (2015)
  arXiv:1506.01415 [hep-ph].

\bibitem{Chien:2015ctp} 
  Y.~T.~Chien, Z.~B.~Kang, F.~Ringer, I.~Vitev and H.~Xing,
  JHEP {\bf 1605}, 125 (2016)
  arXiv:1512.06851 [hep-ph].

\bibitem{Dai:2016hzf} 
  L.~Dai, C.~Kim and A.~K.~Leibovich,
  arXiv:1606.07411 [hep-ph].

\bibitem{Chatrchyan:2012gw} 
  S.~Chatrchyan {\it et al.} [CMS Collaboration],
  JHEP {\bf 1210}, 087 (2012)
  arXiv:1205.5872 [nucl-ex].

\bibitem{Aad:2013ysa} 
  G.~Aad {\it et al.} [ATLAS Collaboration],
  JHEP {\bf 1307}, 032 (2013)
  arXiv:1304.7098 [hep-ex].

\bibitem{ATLAS:2015mla}
 ATLAS Collaboration, T. A. collaboration, 2015. ATLAS-CONF-2015-022.

  
  \bibitem{Bauer:2000yr} 
  C.~W.~Bauer, S.~Fleming, D.~Pirjol and I.~W.~Stewart,
  Phys.\ Rev.\ D {\bf 63}, 114020 (2001),
  [hep-ph/0011336].

\bibitem{Beneke:2002ph} 
  M.~Beneke, A.~P.~Chapovsky, M.~Diehl and T.~Feldmann,
  Nucl.\ Phys.\ B {\bf 643}, 431 (2002)
  doi:10.1016/S0550-3213(02)00687-9
  [hep-ph/0206152].

  
  
\bibitem{Bauer:2001yt} 
  C.~W.~Bauer, D.~Pirjol and I.~W.~Stewart,
  Phys.\ Rev.\ D {\bf 65}, 054022 (2002),
  [hep-ph/0109045].

\bibitem{Aad:2012vca} 
  G.~Aad {\it et al.} [ATLAS Collaboration],
  Phys.\ Lett.\ B {\bf 719}, 220 (2013)
  [arXiv:1208.1967 [hep-ex]].

\bibitem{Abelev:2013kqa} 
  B.~Abelev {\it et al.} [ALICE Collaboration],
  JHEP {\bf 1403}, 013 (2014)
  [arXiv:1311.0633 [nucl-ex]].

\bibitem{Jager:2002xm} 
  B.~Jager, A.~Schafer, M.~Stratmann and W.~Vogelsang,
  Phys.\ Rev.\ D {\bf 67}, 054005 (2003),
  [hep-ph/0211007].

\bibitem{Ellis:2010rwa} 
  S.~D.~Ellis, C.~K.~Vermilion, J.~R.~Walsh, A.~Hornig and C.~Lee,
  JHEP {\bf 1011}, 101 (2010),
  arXiv:1001.0014 [hep-ph].

\bibitem{Kolodrubetz:2016dzb} 
  D.~W.~Kolodrubetz, P.~Pietrulewicz, I.~W.~Stewart, F.~J.~Tackmann and W.~J.~Waalewijn,
  arXiv:1605.08038 [hep-ph].

\bibitem{Vogt:2004ns} 
  A.~Vogt,
  Comput.\ Phys.\ Commun.\  {\bf 170}, 65 (2005)
  [hep-ph/0408244].

\bibitem{Anderle:2015lqa} 
  D.~P.~Anderle, F.~Ringer and M.~Stratmann,
  Phys.\ Rev.\ D {\bf 92}, no. 11, 114017 (2015),
  arXiv:1510.05845 [hep-ph].

\bibitem{Bodwin:2015iua} 
  G.~T.~Bodwin, K.~T.~Chao, H.~S.~Chung, U.~R.~Kim, J.~Lee and Y.~Q.~Ma,
  Phys.\ Rev.\ D {\bf 93}, no. 3, 034041 (2016),
  arXiv:1509.07904 [hep-ph].

\bibitem{Fickinger:2016rfd} 
  M.~Fickinger, S.~Fleming, C.~Kim and E.~Mereghetti,
  arXiv:1606.07737 [hep-ph].

\bibitem{Nadolsky:2008zw} 
  P.~M.~Nadolsky, H.~L.~Lai, Q.~H.~Cao, J.~Huston, J.~Pumplin, D.~Stump, W.~K.~Tung and C.-P.~Yuan,
  Phys.\ Rev.\ D {\bf 78}, 013004 (2008),
  arXiv:0802.0007 [hep-ph].

\bibitem{deFlorian:2007ekg} 
  D.~de Florian, R.~Sassot and M.~Stratmann,
  Phys.\ Rev.\ D {\bf 76}, 074033 (2007)
  arXiv:0707.1506 [hep-ph].

\bibitem{Chien:2015hda} 
  Y.~T.~Chien and I.~Vitev,
  JHEP {\bf 1605}, 023 (2016)
  [arXiv:1509.07257 [hep-ph]].

  
  
\bibitem{Chien:2016led} 
  Y.~T.~Chien and I.~Vitev,
  arXiv:1608.07283 [hep-ph].


\bibitem{Ovanesyan:2011kn} 
  G.~Ovanesyan and I.~Vitev,
  Phys.\ Lett.\ B {\bf 706}, 371 (2012)
  doi:10.1016/j.physletb.2011.11.040
  [arXiv:1109.5619 [hep-ph]].
  


\end{thebibliography}
\end{document}